\documentclass{emulateapj}

\usepackage{epsfig,graphicx}
\newcommand{\oversim}[2]{\protect{\mbox{\lower0.5ex\vbox{%
   \baselineskip=0pt\lineskip=0.2ex
   \ialign{$\mathsurround=0pt #1\hfil##\hfil$\crcr#2\crcr\sim\crcr}}}}} 

\newcommand{\simless} {\mbox{$\,\mathrel{\mathpalette\oversim<}\,$}} 

\shorttitle{Unorthodox Orbits of Substructure Halos}
\shortauthors{Ludlow et al.}

\begin{document}

\title{The Unorthodox Orbits of Substructure Halos} 
\author{
Aaron D. Ludlow\altaffilmark{1}, Julio
F. Navarro\altaffilmark{1,2}, Volker
Springel\altaffilmark{3}, Adrian Jenkins\altaffilmark{4}, \\ Carlos
S. Frenk\altaffilmark{4}, Amina Helmi\altaffilmark{5}}
\affil{$^1$ Department of Physics and Astronomy, University of
Victoria, 3800 Finnerty Rd., Victoria, BC, V8P 5C2, Canada}
\affil{$^2$ Astronomy Department, University of Massachusetts, 710
N. Pleasant St., Amherst, MA 01003, USA}
\affil{$^3$ Max-Planck Institute for Astrophysics, Karl-Schwarzschild Str. 1, D-85748,
Garching, Germany} 
\affil{$^4$ Institute of Computational Cosmology,
Department of Physics, University of Durham, Science Laboratories,
South Road, Durham, DH1 3LE, UK}
\affil{$^5$ Kapteyn Institute, P.O. Box 800, 9700 AV Groningen, The Netherlands}


\begin{abstract}
We use a suite of cosmological N-body simulations to study the
properties of substructure halos (subhalos) in galaxy-sized cold dark
matter halos. We extend prior work on the subject by considering the
whole population of subhalos {\it physically associated} with the main
system. These are defined as subhalos that have at some time in the
past been within the virial radius of the halo's main progenitor and
that have survived as self-bound entities to $z=0$.  We find that this
population extends beyond {\it three times} the virial radius, and
contains objects on extreme orbits, including a few with velocities
approaching the nominal escape speed from the system. We trace the
origin of these unorthodox orbits to the tidal dissociation of bound
groups of subhalos, which results in the ejection of some subhalos
along tidal streams. Ejected subhalos are primarily low-mass systems,
leading to mass-dependent biases in their spatial distribution and
kinematics: the lower the subhalo mass at accretion time, the less
centrally concentrated and kinematically hotter their descendant
population. The bias is strongest amongst the most massive subhalos,
but disappears at the low-mass end: below a certain mass, subhalos
behave just like test particles in the potential of the main halo.
Overall, our findings imply that subhalos identified within the virial
radius represent a rather incomplete census of the substructure
physically related to a halo: only about {\it one half} of all
associated subhalos are found today within the virial radius of a
halo, and many relatively isolated halos may have actually been
ejected in the past from more massive systems. These results may
explain the age dependence of the clustering of low-mass halos
reported recently by Gao et al, and has further implications for (i)
the interpretation of the structural parameters and assembly histories
of halos neighboring massive systems; (ii) the existence of low-mass
dynamical outliers, such as Leo I and And XII in the Local Group; and
(iii) the presence of evidence for evolutionary effects, such as tidal
truncation or ram-pressure stripping, well outside the traditional
virial boundary of a galaxy system.

\end{abstract}


\section{Introduction}\label{sec:intro}

In the current paradigm of structure formation, the concordance $\Lambda$CDM scenario, the dark matter halos that host galaxy systems are assembled hierarchically, through the merger and accretion of smaller subunits. One relic of this process is the presence of {\it substructure}, which consists of the self-bound cores of accreted subsystems that have so far escaped full disruption in the tidal field of the main halo (Klypin et al 1999; Moore et al 1999).

Although substructure halos (referred to hereafter as ``subhalos'',
for short) typically make up only a small fraction ($5$ to $10\%$) of
the total mass of the system, they chart the innermost regions of
accreted subsystems, and are thus appealing tracers of the location
and kinematics of the {\it galaxies} that subhalos may have
hosted. Substructure is thus a valuable tool for studying galaxies
embedded in the potential of a much larger system, such as satellite
galaxies orbiting a primary, or individual galaxies orbiting within a
group or cluster of galaxies.

This realization has prompted a number of studies over the past few years, both analytical and numerical, aimed at characterizing the main properties of subhalos, such as their mass function, spatial distribution, and kinematics (e.g. Ghigna et al 1998, 1999; Moore et al 1999; Taylor \& Babul 2005a,b; Benson 2005; Gao et al 2004; Diemand et al 2007a,b).

Consensus has been slowly but steadily emerging on these issues. For
example, the {\it mass function} of subhalos has been found to be
rather steep, $dN/dM \propto M_{\rm sub}^{-1.9}$ or steeper, implying
that the subhalo population is dominated in number by low-mass systems
but that most of the substructure mass resides with the few most
massive subhalos (Springel et al 2001, Helmi, White \& Springel 2002,
Gao et al 2004). Confirmation of this comes from the fact that the
total fraction of mass in subhalos is rather low (typically below
10\%) even in the highest resolution simulations published so far
(although see Diemand et al 2007a,b for a differing view).

Subhalos have also been found to be {\it spatially biased} relative to
the smooth dark matter halo where they are embedded, avoiding in
general the innermost regions. Furthermore, the number density profile
of the subhalo population also differs markedly from that of galaxies
in clusters, and possibly from the radial distribution of luminous
satellites around the Milky Way (Kravstov et al 2004; Willman et al
2004; Madau et al 2008). This precludes identifying directly the
population of ``surviving'' subhalos with galaxies in clusters, and
highlights the need for either more sophisticated numerical modeling
techniques, or for pairing up the N-body results with semi-analytic
modeling in order to trace more faithfully the galaxy population
(Springel et al 2001, De Lucia et al 2004, Gao et al 2004, Croton et
al 2006).


One intriguing result of all these studies has been the remarkably weak dependence of the properties of substructure on subhalo mass. Gao et al (2004) and Diemand, Moore \& Stadel (2004), for example, find that the radial distribution of subhalos is largely independent of their self-bound mass. This is surprising given the strong mass dependence expected for the processes that dictate the evolution of subhalos within the main halo, such as dynamical friction and tidal stripping. Although efficient mixing within the potential of the main halo is a possibility, an alternative explanation has been advanced by Kravtsov, Gnedin \& Klypin (2004).

These authors argue that the {\it present-day} mass of a subhalo may be a poor indicator of the {\it original} mass of the system, which may have been substantially larger at the time of accretion, and used this idea to motivate how the faintest dwarf companions of the Milky Way were able to build up a sizable stellar mass through several episodes of star formation despite their shallow present-day potential wells. The same idea was also adopted by Libeskind et al (2007) as a possible reason for the peculiar spatial alignment of satellites around the Milky Way (Lynden-Bell 1976, 1982; Kunkel \& Demers 1976; Kroupa, Thies \& Boily 2005).

We revisit here these issues with the aid of a suite of
high-resolution N-body simulations of galaxy-sized halos. We extend
prior work by carefully tracking the orbits of surviving subhalos back
in time. This allows us to select a complete set of subhalos {\it
physically associated} with the main halo, rather than only the ones
that happen to be within the virial radius at a particular time. As we
discuss below, a large fraction of the associated subhalo population
are on unorthodox orbits that take them well beyond the virial radius,
a result with important implications for studies of satellite galaxies
and of halos clustered around massive systems.

The plan of this paper is as follows. We introduce briefly the
numerical simulations in \S~\ref{sec:numexp} and describe our subhalo
detection algorithm and tracking method in \S~\ref{sec:anal}. Our main
results are presented in \S~\ref{sec:res}: we begin by exploring the
subhalo spatial distribution and kinematics, as well as their
dependence on mass, and discuss the consequences of our findings for
the subhalo mass function.  We end with a brief summary and discussion
of possible implications and future work in \S~\ref{sec:conc}.

\section{The Numerical Simulations}
\label{sec:numexp}

\subsection{The Cosmological Model}

All simulations reported here adopt the concordance $\Lambda$CDM
model, with parameters chosen to match the combined analysis of the
first-year WMAP data release (Spergel et al 2003) and the 2dF Galaxy
Redshift Survey (Colless et al 2001). The chosen cosmological
parameters are $\Omega_{\rm m}=\Omega_{\rm dm}+\Omega_{\rm b}=0.25$,
$\Omega_{\rm b}=0.045$, $h=0.73$, $\Omega_{\rm \Lambda} = 0.75$,
$n=1$, and $\sigma_8=0.9$. Here $\Omega$ denotes the present-day
contribution of each component to the matter-energy density of the
Universe, expressed in units of the critical density for closure,
$\rho_{\rm crit}=3H^2/8\pi G$; $n$ is the spectral index of the
primordial density fluctuations, and $\sigma_8$ is the rms linear mass
fluctuation in spheres of radius $8 \, h^{-1}$ Mpc at $z=0$. Hubble's
``constant'' is given by $H(z)$ and parameterized at $z=0$ by
$H(z=0)=H_0=100\, h $ km s$^{-1}$ Mpc$^{-1}$.

\subsection{The Runs}

Our analysis is based on a suite of $5$ high-resolution simulations of the formation of galaxy-sized $\Lambda$CDM halos. The simulations target halos of virial mass\footnote{We define the virial mass of a halo, $M_{200}$, as that contained within a sphere of mean density $200\times \rho_{\rm crit}$. The virial mass defines implicitly the virial radius, $r_{200}$, and virial velocity, $V_{200}=\sqrt{GM_{200}/r_{200}}$, of a halo, respectively. We note that other definitions of ``virial radius'' have been used in the literature; the most popular of the alternatives adopts a density contrast (relative to critical) of $\Delta\approx 178 \, \Omega_{\rm m}^{0.45}\sim 100$ (for our adopted cosmological parameters, see Eke et al 1996). We shall refer to these alternative choices, where appropriate, with a subscript indicating the value of $\Delta$; i.e., $r_{100}$ is the virial radius obtained assuming $\Delta=100$.},
$M_{200} \sim 10^{12}\, h^{-1} \, M_{\odot}$, and have at $z=0$ between $3$ and $5$ million particles within the virial radius, $r_{200}$. Each halo was selected at random from a list of candidates compiled from a cosmological N-body simulation of a large ($100\, h^{-1}$ Mpc) periodic box and resimulated individually at higher resolution using the technique described in detail by Power et al (2003). We imposed a mild isolation criterion (that no neighbors with mass exceeding $5\times 10^{11} h^{-1} M_\odot$ be found within $1 h^{-1}$ Mpc at $z=0$) in order to exclude systems formed in the periphery of much larger groups or clusters.

The simulations were run with {\tt Gadget2}, a massively-parallel
cosmological N-body code (Springel 2005). Particle pairwise
interactions were softened using the ``optimal'' gravitational
softening length scale suggested by Power et al (2003); i.e., a spline
lengthscale $h_{\rm s}=1.4 \epsilon_{\rm G} \approx 4\,
r_{200}/\sqrt{N_{200}}$, kept fixed in comoving coordinates. Numerical details of each run are listed in
Table~\ref{tab:numexp}.

\section{The Analysis}
\label{sec:anal}

\subsection{Substructure Finding}
\label{ssec:subfind}

We use {\tt SUBFIND} (Springel et al 2001) in order to identify
self-bound structures in N-body simulations. {\tt SUBFIND} finds
substructure within friends-of-friends (FOF; Davis et al 1985)
associations by locating overdense regions within each FOF halo and
identifying the bound subset of particles associated with each
overdensity. {\tt SUBFIND} also works recursively and its output
readily identifies ``subhalos within subhalos'', thus characterizing
fully the various levels of the hierarchy of substructure present
within a given FOF halo. We retain for our catalogue all {\tt SUBFIND}
subhalos with more than $20$ particles.

The main output of {\tt SUBFIND} is a list of subhalos within each FOF
halo, together with their structural properties. For the purposes of
this paper, we shall focus on: (i) the subhalo self-bound mass,
$M_{\rm sub}$; (ii) the peak of its circular velocity profile
(characterized by $r_{\rm max}$ and $V_{\rm max}$); and (iii) the
position of the subhalo center, which we identify with the particle
with minimum gravitational potential energy. We have run {\tt SUBFIND}
on all $100$ snapshots (equally spaced in scale factor, $a$) of each
of our runs, and are therefore able to track in detail the evolution
of individual subsystems and their particle members.

\subsection{Substructure Tracking}
\label{ssec:subtrack}

Our analysis focuses on all {\it surviving} subhalos at $z=0$ and
relies heavily on tracking accurately their accretion history. To this
aim, we trace each subhalo backwards in time by identifying the
central particle at $z=0$ and searching for the group it belongs to in
the immediately preceding snapshot. A new central particle is then
selected and the procedure is iterated backwards in time until $z=9$,
the earliest time we consider in the analysis.

This procedure leads in general to a well-defined evolutionary track
for each subhalo identified at $z=0$. When no subhalo is found to
contain a subhalo's central particle in the immediately preceding
snapshot, the search is continued at earlier times until either a
progenitor subhalo is found or $z=9$ is reached. This is necessary
because a subhalo may temporarily disappear from the catalogue,
typically at times when it falls below the minimum particle number or
else when it is passing close to the center of a more massive system
and its density contrast is too low to be recognized by {\tt
SUBFIND}. Our procedure overcomes this difficulty and in most cases
recovers the subhalo at an earlier time. We note that these
complications are a fairly common occurrence in the analysis
procedure, and we have gone to great lengths to make sure that these
instances are properly identified and dealt with when constructing our
subhalo catalogue and their accretion histories.

The tracking procedure described above defines a unique trajectory for
each subhalo identified at $z=0$. This trajectory may be used to
verify whether a subhalo has, at any time in the past, been accreted
within the (evolving) virial radius of the main halo. If this is the
case, we record the time it first crosses $r_{200}(z)$ as the
``accretion redshift'', $z_{\rm acc}$, and label the subhalo as {\it
associated} with the main system. Analogously, we identify a set of
{\it associated dark matter} particles by compiling a list of all
particles that were at some time within the virial radius of the main
halo but are not attached to any substructure at $z=0$. On the other
hand, halos that have {\it never} been inside the virial radius of the
main halo will be referred to as ``field'' or ``infalling'' halos.

Using the subhalo trajectories, we compute and record a few further quantities of interest for each subhalo; namely,
\begin{itemize}

\item its ``turnaround'' distance, $r_{\rm ta}$, defined as the {\it maximum separation} between a subhalo and the center of the main progenitor before $z=z_{\rm acc}$ (for associated subhalos) or before $z=0$ (for field subhalos); 

\item the structural properties of associated subhalos at $z=0$ and at accretion time, such as mass and peak circular velocity;

\item an apocentric distance, $r_{\rm apo}$, defined as the apocenter  of its orbit computed at $z=0$ using the subhalo's instantaneous  kinetic energy and orbital angular momentum, together with the  potential of the main halo{\footnote{We have checked that our results are insensitive to the triaxial nature of the halo by recomputing $r_{\rm apo}$ using the potential along each of the principal axes of the halo's mass distribution rather than the spherical average. This leads to typical variations of less then $\sim 20\%$ in $r_{\rm apo}$.}}

\end{itemize}

Subhalo quantities measured at accretion time will be referred to by
using the sub/superscript ``{\rm acc}''; for example, $V_{\rm
max}^{\rm acc}$ refers to the peak circular velocity of a subhalo at
$z=z_{\rm acc}$. Quantities quoted without superscript are assumed to
be measured at $z=0$ unless otherwise specified; e.g., $V_{\rm
max}=V_{\rm max}(z=0)$.

\begin{figure} [ht]
\begin{center}
\epsfig{file=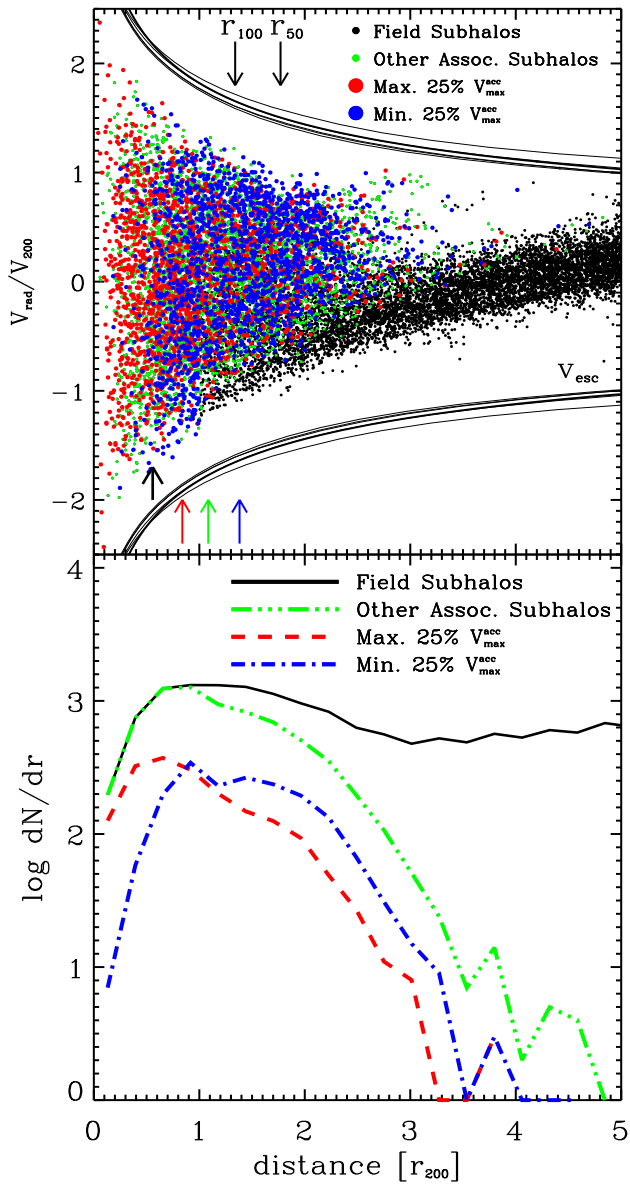,height=16.5 cm,width=8.5 cm,angle=0}
\caption[] 
{\footnotesize{{{\it Upper panel:} Radial velocity versus distance to
the main halo center for all subhalos within $5\times r_{200}$ in our
simulations. Velocities and distances are normalized to the virial
velocity, $V_{200}$, and virial radius, $r_{200}$, of each host. All
``associated'' halos are shown in color, subhalos on first infall are
shown in black. Different colors are used according to the peak
circular velocity of the subhalo at the time of accretion. Blue
denotes the quartile with smallest $V_{\rm max}^{\rm acc}$, red those
with largest $V_{\rm max}^{\rm acc}$. Green denotes the rest of the
associated subhalo population.  Upward vertical arrows of matching
color indicate the half-number radius for the various subhalo
populations. A shorter black arrow marks the half-number radius for
``associated'' dark matter particles. We find that $65\%$ of subhalos
in the range $r_{200} < r < 2 \, r_{200}$ are actually ``associated''
and have thus already been within the host virial radius in the
past. Roughly one third of subhalos between $2 \, r_{200} < r < 3 \,
r_{200}$ are also physically ``associated'' with the main halo. The
upper and lower bounding curves denote the escape velocity for each of
the five simulated halos. {\it Lower panel:} Radial distribution of
subhalos. Color key is the same as in the upper panel.}}}
\label{fig:VradVersusRad}
\end{center}
\end{figure}

\section{Results}
\label{sec:res}

The basic properties of our simulated halos at $z=0$ are presented in
Table~\ref{tab:numexp}. Here, $\epsilon_{\rm G}(=h_{\rm s}/1.4)$ is
the {\tt Gadget} gravitational softening input parameter, and
$M_{200}$, $r_{200}$, and $N_{200}$ are, respectively, the halo virial
mass, radius, and number of particles within
$r_{200}$. Table~\ref{tab:numexp} also lists the peak of the circular
velocity of the main halo, $V_{\rm max}$, and its location, $r_{\rm
max}$; the total number of ``associated'' subhalos; as well as the
number of those found within various characteristic radii at $z=0$.

\begin{figure} [ht]
\begin{center}
\epsfig{file=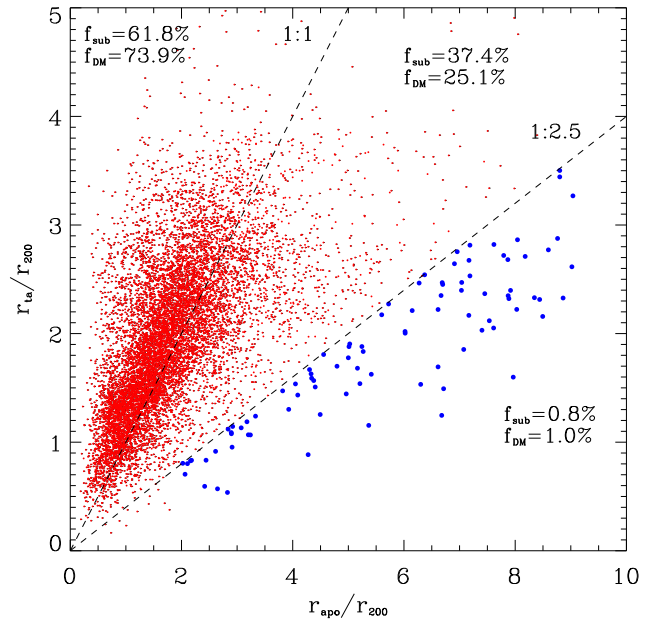,height=8.5 cm,width=8.5 cm,angle=0}
\caption[] 
{\footnotesize{{Turnaround radius versus apocentric distance at $z=0$
for all {\it associated } subhalos in our simulations. The turnaround
distance is the maximum distance from the main halo before
accretion. Subhalos on ``traditional'' orbits are expected to have
$r_{\rm apo} < r_{\rm ta}$ and, thus, to be to the left of the $1$:$1$
curve in this plot. Subhalos near the $1$:$1$ line have $r_{\rm apo}
\approx r_{\rm ta}$ and are therefore on orbits which have not been
decelerated substantially since turnaround. Subhalos with $r_{\rm apo}
> r_{\rm ta}$ are on unorthodox orbits and they have {\it gained}
orbital energy during or after accretion. The blue symbols in the
panel highlight subhalos on extreme orbits, that will take them more
than $\sim 2.5$ times farther than their turnaround radius. The
fraction of associated subhalos and associated dark matter particles
in each region of the plot is given in the legends.}}}
\label{fig:RapoRturn}
\end{center}
\end{figure}

\subsection{Subhalos beyond the virial radius}
\label{ssec:outersubh}

One surprise in Table~\ref{tab:numexp} is that the number of
``associated'' subhalos exceeds by about a factor of $\sim 2$ the
total number of subhalos identified within $r_{200}$. This result is
also illustrated in Fig.~\ref{fig:VradVersusRad}, where we show, at
$z=0$, the distance from the main halo center vs radial velocity for
all subhalos identified in our simulations. Distances and velocities
have been scaled to the virial quantities of each primary
halo. Colored dots are used to denote ``associated'' subhalos, black
symbols for ``field'' halos. Different colors correspond to different
subhalo masses, as measured by the peak circular velocity at accretion
time (in units of the present-day primary halo virial velocity,
$V_{200}$): red is used for subhalos with $V_{\rm max}^{\rm acc}\ge
0.72 \, V_{200}$, blue for those with $V_{\rm max}^{\rm acc}\le 0.038
\, V_{200}$, green for the rest.

Note that the distribution of associated subhalos extends well past $\sim 3\, r_{200}$; indeed, a few associated subsystems are found at $r\sim 4\, r_{200}$ moving outwards with radial velocity of order $V_r \sim V_{200}$. A careful search shows that there are actually several associated subhalos presently at distances larger than $\sim 5 \, r_{200}$.

This result is unexpected in simple formation scenarios, such as the {\it spherical secondary infall model} (SSIM, for short). SSIM identifies at any time three distinct regions around a halo: (i) an inner ``virialized'' region where accreted mass shells have had time to cross their orbital paths; (ii) a surrounding ``infall'' region, where shells are still on first approach and have not yet crossed; and (iii) a still expanding outer envelope beyond the current turnaround radius (Fillmore \& Goldreich 1984; Bertschinger 1985, White et al 1993, Navarro \& White 1993).

One of the premises of the secondary infall model is that the energy of a mass shell accreted into the halo is gradually reduced after its first pericentric passage until it reaches equilibrium. During this process, the apocentric distance of the shell is constantly reduced; for example, taking as a guide the SSIM self-similar solutions of Bertschinger (1985), the {\it second} apocenter of an accreted shell (the first would be its ``turnaround'' radius) is roughly $90\%$ of its turnaround distance, $r_{\rm ta}$, and the shell gradually settles to equilibrium, approaching a periodic orbit with $r_{\rm apo}\sim 0.8 \, r_{\rm ta}$. Thus, according to the SSIM, few, if any, associated subhalos are expected to populate the region outside $\sim 0.8 \, r_{\rm ta}\approx 1.6 \, r_{200}$. This is clearly at odds with the results shown in Fig.~\ref{fig:VradVersusRad} and Table~\ref{tab:numexp}: more than a {\it quarter} of all associated subhalos are found beyond $1.6 \ r_{200}$ at $z=0$!

\begin{figure*} [ht]
\begin{center}
\epsfig{file=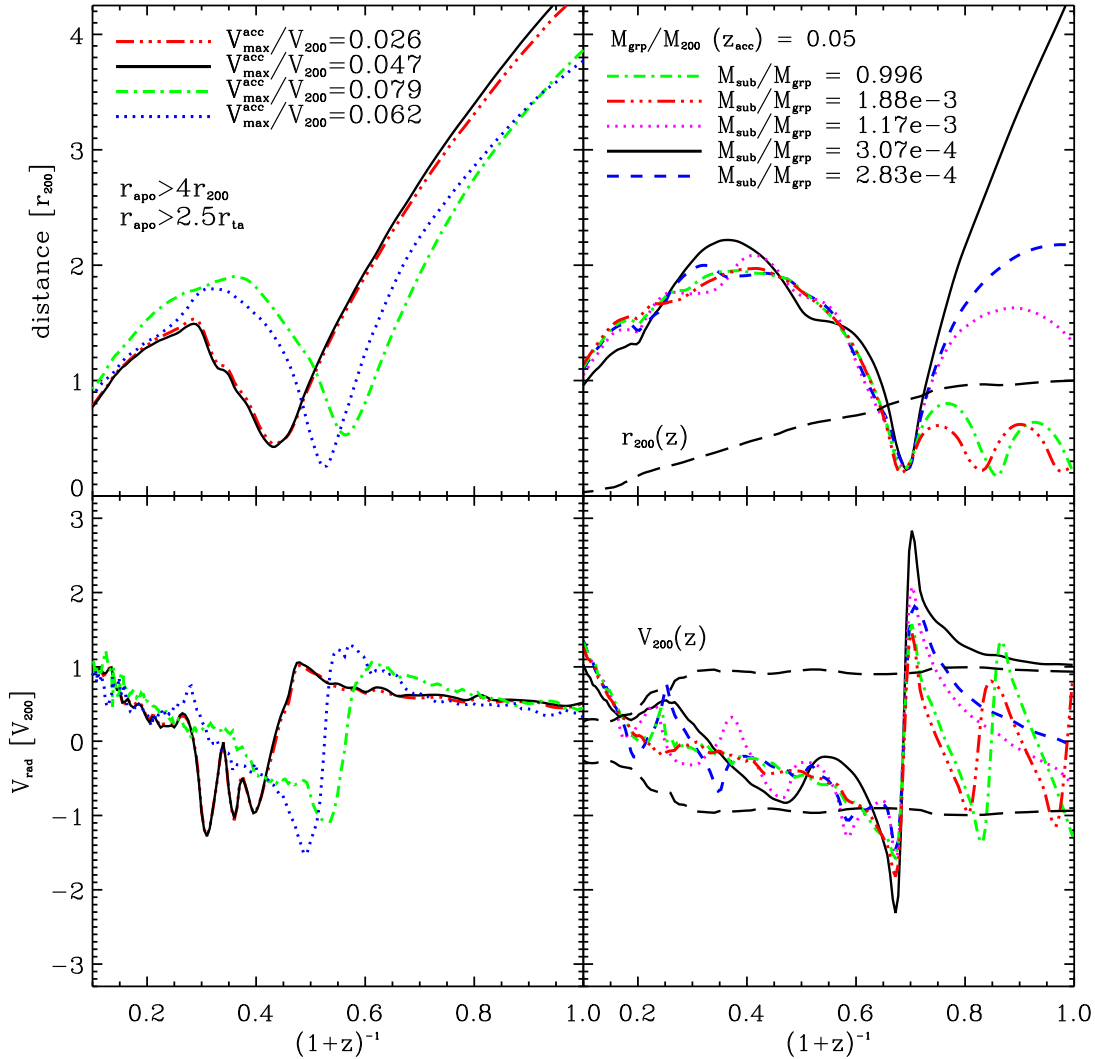,height=15. cm,width=15. cm,angle=0}
\caption[] 
{\footnotesize{{ Orbital trajectories of selected subhalos. Upper
panels show the distance to the center of the main progenitor as a
function of expansion factor. The top-left panel shows the
trajectories of $4$ subhalos on ``extreme'' orbits (blue points in
Fig.~\ref{fig:RapoRturn}). Note that all of these systems gain energy
during their first pericentric approach to the main halo. The
top-right panel illustrates that interactions occurring
during the tidal dissociation of bound groups of subhalos are
responsible for propelling some satellites onto extreme orbits. At
pericentric approach, the tidal field of the main halo breaks apart
the group, and redistributes each member onto orbits of varying
energy. The most affected are, on average, the least massive members
of the group, some of which are pushed onto orbits with extremely
large apocenters. The dashed curve shows the growth of the virial
radius of the most massive progenitor of the main halo. Bottom panels
show the radial velocity of the subhalos shown in the upper panels.}}}
\label{fig:orbits}
\end{center}
\end{figure*}

\begin{figure} []
\begin{center}
\includegraphics[width=\linewidth,clip]{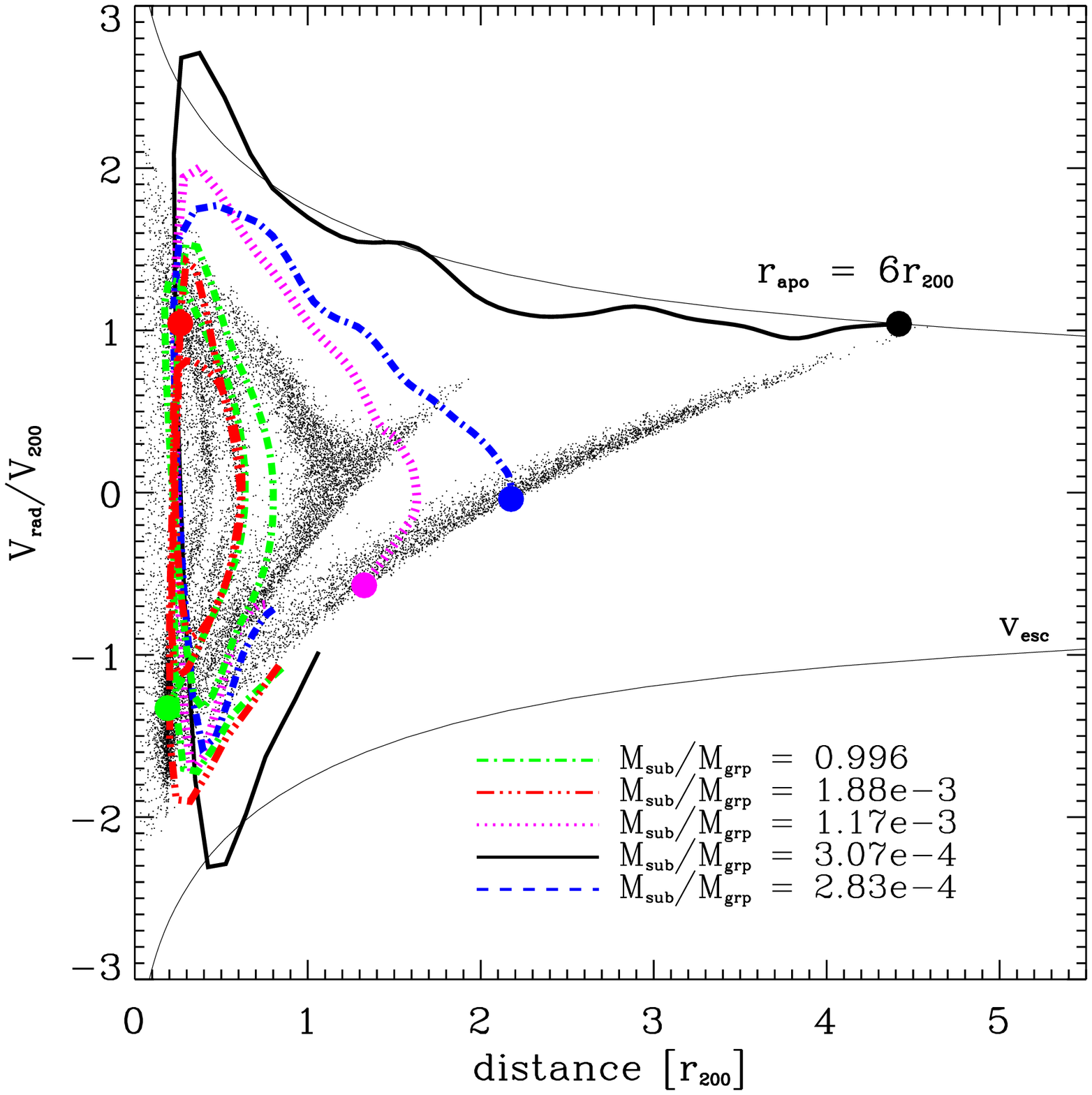}
\caption[] 
{\footnotesize{{Black dots show the position, at $z=0$, of particles
belonging, at accretion time, to the most massive subhalo of the group
shown in the right-hand panels of Fig.~\ref{fig:orbits}. This subhalo
has been fully disrupted in the potential of the main halo. Large
circles show the position of the center of mass of the other
(surviving) subhalos in the group. Curves show the evolution of each
of these subgroups since accretion. Note how the surviving subhalos
align themselves with tidal streams stripped from the main subhalo
during the disruption process. The ``ejection'' of subhalos is thus due to
the same mechanism that leads to the formation of outgoing tidal tails
in a merger event and should occur naturally during the tidal
dissociation of any bound group of subhalos.}}}
\label{fig:PhaseAcc}
\end{center}
\end{figure}

\subsection{The orbits of associated subhalos}
\label{ssec:subhorbs}
The discrepancy between the simulation results and the naive
expectation of the SSIM was pointed out by Balogh, Navarro \& Morris
(2000), and confirmed by subsequent studies (Mamon et al 2004, Gill et
al 2005, Diemand et al 2007) but its physical origin has not yet been
conclusively pinned down. Associated subhalos found today beyond their
turnaround radius have clearly evolved differently from the SSIM
prescription, and it is instructive to study the way in which the
difference arises.

One possibility is that deviations from spherical symmetry during
accretion might be responsible for the outlying associated
subhalos. Accretion through the filamentary structure of the cosmic
web surrounding the halo, for example, might result in a number of
subhalos on orbits of large impact parameter that simply ``graze'' the
main halo and are therefore not decelerated significantly after their
first pericentric approach, as assumed in the secondary infall
model. These subhalos would lose little orbital energy, and should
presumably be today on orbits with apocentric distances of the order
of their original turnaround radii. According to the analytic
calculation of Mamon et al (2004), systems on such orbits may reach
distances as far as $r_{\rm apo} \sim 2.3\, r_{200}$.

We explore this in Fig.~\ref{fig:RapoRturn}, where we show the turnaround radius of each associated subhalo versus their apocentric distance estimated at $z=0$, both normalized to the virial radius of the main halo. Subhalos that have followed the traditional orbits expected from the SSIM should lie to the left of the $1$:$1$ curve in this panel. These, indeed, make up the bulk ($\sim 62\%$) of the associated population.

Note as well that there are a number of subhalos near the $1$:$1$ line, whose orbits have not been decelerated since accretion into the main halo. These are objects that are either on their way to first pericentric passage or, as discussed in the above paragraph, that have somehow evaded significant braking during accretion.

More intriguingly, Fig.~\ref{fig:RapoRturn} also shows that there are
a significant number of subhalos on decidedly unorthodox orbits, with
apocenters exceeding their SSIM theoretical ``maximum''; i.e., $r_{\rm
apo} > r_{\rm ta}$. Indeed, $\sim 38\%$ of associated subhalos are on
such orbits, and about $\sim 1\%$ are on orbits so extreme that their
apocenters exceed their original turnaround distance by more than a
factor of $\sim 2.5$ (the latter are highlighted in blue in
Fig.~\ref{fig:RapoRturn} if, in addition, $r_{\rm apo} > 2\,
r_{200}$). The large fraction of systems in such peculiar orbits,
where the subhalo has {\it gained} orbital energy since turnaround,
indicates that deviations from spherical symmetry play a minor role in
pushing subhalos beyond $r_{200}$, and suggests that another mechanism
is at work.

\subsection{Subhalo mass dependence of unorthodox orbits}
\label{ssec:subhmdep}

One clue to the mechanism responsible for pushing some subhalos onto
highly energetic orbits is the dependence of the effect on the mass of
the subhalo. This is illustrated in the bottom panel of
Fig.~\ref{fig:VradVersusRad}, which shows that low-mass subhalos are
the ones being preferentially pushed to the outskirts of the halo.

Further clues result from inspecting individually the trajectories of
some of the subhalos on extreme orbits. This is shown in the top-left
panel of Fig.~\ref{fig:orbits}, where we show the orbits of a few of
the associated subhalos with $r_{\rm apo} > 2.5 \, r_{\rm
ta}$. Interestingly, all of these subhalos have very low mass at
accretion ($V_{\rm max}^{\rm acc} \simless 0.08 V_{200}$) and acquire
their ``boost'' in orbital energy during their {\it first} pericentric
passage.

The ``wiggles'' in their trajectories prior to pericenter betray the
fact that they actually belong to a bound system of multiple subhalos
accreted as a single unit (Sales et al 2007a, Li \& Helmi 2007). This
is shown more clearly in the top-right panel of Fig.~\ref{fig:orbits},
where we show the trajectories of $5$ subhalos belonging to one such
group. The mass of the group is concentrated in the most massive
member (see legends in the figure), which is surrounded (prior to
accretion) by 4 bound satellites. The group contributes about $5\%$ of
the main halo's mass at accretion time, $a_{\rm acc}=(1+z_{\rm
acc})^{-1} \approx 0.65$. The group as a whole turns around at $a_{\rm
ta}\approx 0.35$ and accretes on a ($r_{\rm per}$:$r_{\rm
apo})=(1$:$10)$ orbit that reaches $r_{\rm per} \sim 0.25\, r_{200}$
at $a_{\rm per}\sim 0.69$. Adding to this evidence, we find that $\sim
95\%$ of subhalos with $r_{\rm apo} > 2\ r_{200}$ and $r_{\rm apo} >
2.5 \ r_{\rm ta}$ were each, at accretion, members of an FoF group
with multiple subhalos.

During pericentric passage, the group is dissociated by the tidal
field of the main halo, and its $5$ members are flung onto orbits of
widely different energy. The most massive object (single dot-dashed
curve in the right panels of Fig.~\ref{fig:orbits}) follows a
``traditional'' orbit, rebounding to a second apocenter which is only
$\sim 30\%$ of its turnaround distance. The rest evolve differently;
the least massive subhalos, in particular, tend to {\it gain} energy
during the disruption of the group and recede to a second apocenter
well beyond the original turnaround. As anticipated by the work of
Sales et al (2007b) this is clearly the result of energy
re-distribution during the tidal dissociation of the group.

The bottom panels in Fig.~\ref{fig:orbits}, which show the evolution
of the radial velocity of each subhalo, confirm this suggestion.  The
least massive member of the group is, in this case, the least bound as
well, judging from its excursions about the group's center of
mass. This subhalo (solid black line in the right panels of
Fig.~\ref{fig:orbits}) happens to be approaching the group's orbital
pericenter at about the same time as when the group as a whole
approaches the pericenter of its orbit. This coincidence in orbital
phase allows the subhalo to draw energy from the interaction; the
subhalo is thus propelled onto an orbit that will take it beyond three
times its turnaround distance, or $\sim 6 \, r_{200}$. Although
technically still bound, for all practical purposes this subhalo has
been physically ejected from the system and might be easily confused
for a system that has evolved in isolation.

There are similarities between this ejection process and the findings
of early N-body simulations, which showed that a small but sizable
fraction of particles are generally ejected during the collapse of a
dynamically cold N-body system (see, e.g., van Albada 1982). The
latter occur as small inhomogeneities are amplified by the collapse,
allowing for substantial energy re-distribution between particles as
the inhomogeneities are erased during the virialization of the
system. 

In a similar manner, the tidal dissociation of bound groups of
subhalos leads to the ejection of some of the group members. The main
difference is that, in this case, no major fluctuations in the
gravitational potential of the main system occur. Indeed, in the case
shown in the right-hand panels of Fig.~\ref{fig:orbits} the main halo
adds only $\sim 5\%$ of its current mass and its potential changes
little in the process.

A more intuitive illustration is perhaps provided by
Fig.~\ref{fig:PhaseAcc}, where we show, in the $r$-$V_{\rm rad}$ plane
and at $z=0$, the location of the same accreted group of
subhalos. Black dots indicate particles beloging to the main subhalo
at the time of accretion. Large circles mark the location of the
center of mass of each surviving subhalo, and the curves delineate
their past evolution in the $r$-$V_{\rm rad}$ plane. The three
outermost subhalos track closely a stream of particles formerly
belonging to the main subhalo: the ``ejected'' subhalos are clearly
part of the outgoing ``tidal tail'' stripped from the system during
first approach. The origin of subhalos on extreme orbits is thus the
same as that of particles at the tip of the outgoing tidal tails
during a merger, and it should therefore be a common occurrence during
the accretion of any bound group of subhalos.

\begin{figure} []
\begin{center}
\includegraphics[width=\linewidth,clip]{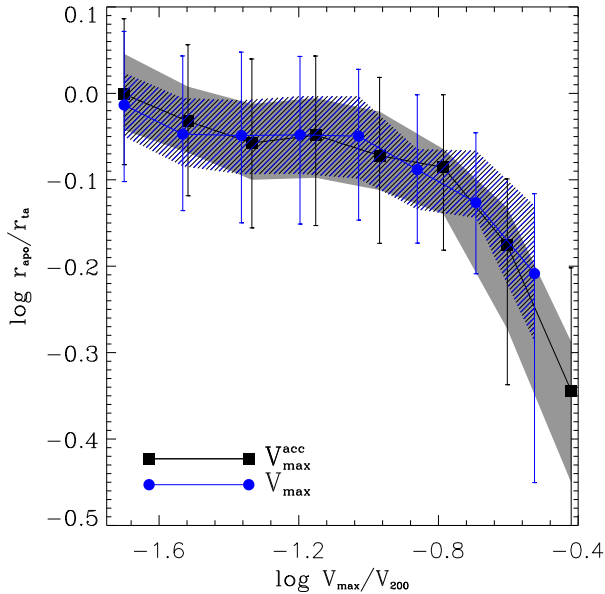}
\caption[] 
{\footnotesize{{The ratio of apocentric radius (estimated at $z=0$) to
turnaround distance as a function of the peak circular velocity,
$V_{\rm max}$, of a subhalo. Two estimates of $V_{\rm max}$ are used
for each subhalo, one measured at accretion time and another at
$z=0$. Symbols correspond to the median of the distribution, shaded
areas encompass 25\% of the distribution around the median, and the
extremes of the error bars correspond to the 25th and 75th
centiles. Note that only fairly massive associated subhalos are today
on orbits substantially more bound than when they turned around. The
median apocentric radius of low-mass subhalos is of order of the
virial radius, indicating that about {\it half} of all associated
subhalos spend a substantial fraction of their orbital period outside
$r_{200}$. Note that the effect depends only weakly on $V_{\rm max}$
below a certain threshold; this presumably indicates that, below a
certain mass limit, subhalos behave like test particles in the
potential of the main halo.}}}
\label{fig:RaRtVmax}
\end{center}
\end{figure}

It is also important to point out that not all low-mass subhalos are
affected equally. For example, despite being of comparable mass to the
ejected object, one of the low mass members of the group ends up on an
orbit almost as tightly bound as the main subhalo (red triple
dot-dashed curve in Fig.~\ref{fig:orbits}). This shows that the
orbital fate of a subhalo is mainly determined by its orbital phase
within the accreting group at the time of accretion. Depending on
this, subhalos may either {\it lose} or {\it gain} orbital energy
during the interaction.

\begin{figure*} [ht]
\begin{center}
\includegraphics[width=\linewidth,clip]{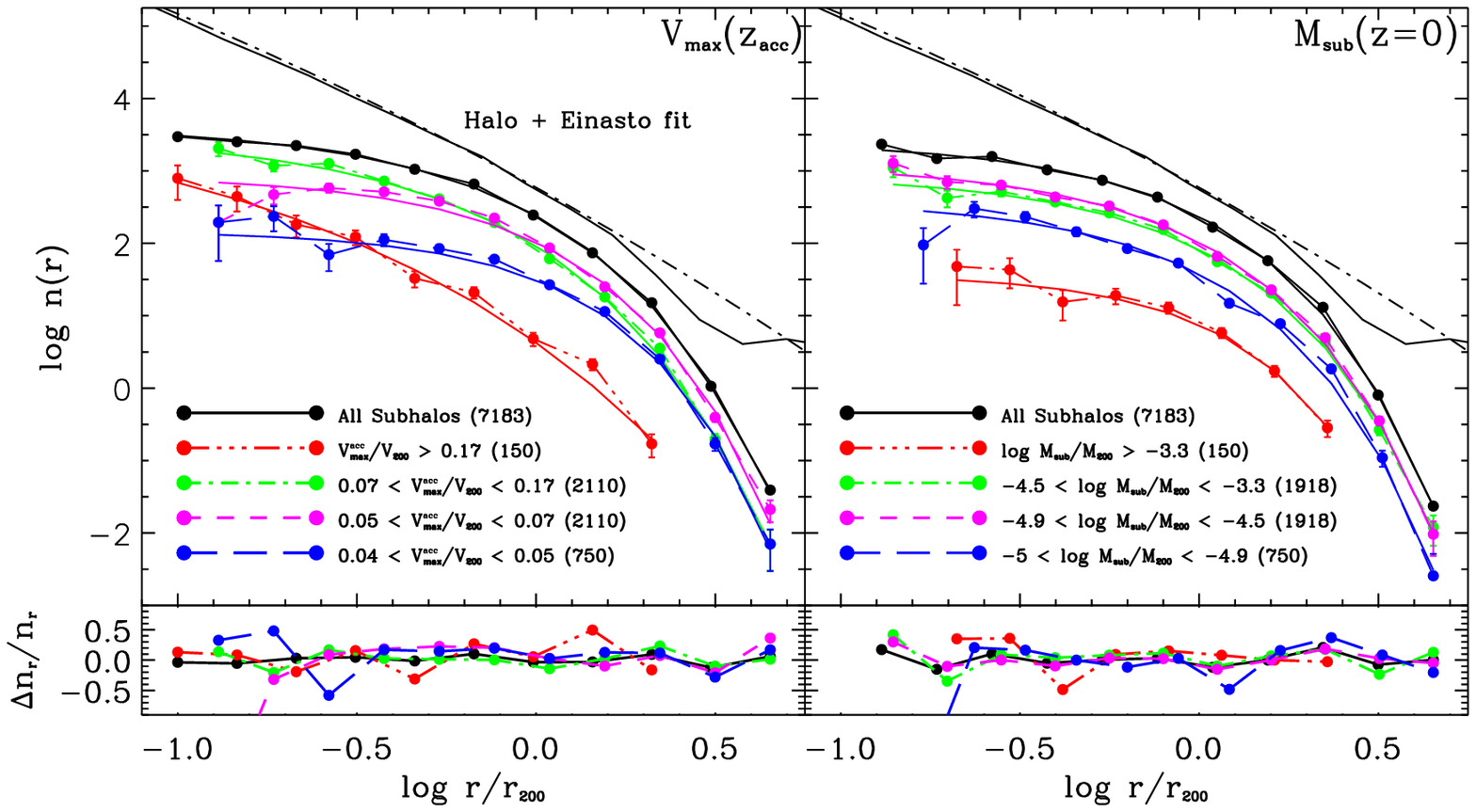}
\caption[] 
{\footnotesize{{Number density profile of associated subhalos, after stacking the results of all $5$ simulations in our series. The black solid symbols show the result for all subhalos, the other symbols correspond to various subsamples obtained after splitting by either $V_{\rm max}^{\rm acc}$ (left panel) or by subhalo mass at $z=0$ (right-hand panel). Details on the velocity and mass range for each subsample are given in the legend. Solid lines through each curve correspond to the best fits obtained with eq.~\ref{eq:rhoalpha}. The parameters of each fit are listed in Table~\ref{tab:nrho}. Lines without symbols show the dark matter density profile. Note that the spatial distribution of subhalos depends sensitively on subhalo mass when measured by $V_{\rm max}^{\rm acc}$, but that, in agreement with prior work, the mass bias essentially disappears when adopting $M_{\rm sub}$ to split the sample. See text for further discussion.}}}
\label{fig:NrhoFit}
\end{center}
\end{figure*}

Low mass halos are, however, the ones preferentially ``ejected'' or
placed on high-energy orbits through this process (see
Fig.~\ref{fig:VradVersusRad} and Fig.~\ref{fig:NrhoFit}).  This is
because low-mass members of accreting groups will have orbits of
larger amplitude about its center of mass, enhancing the probability
of capturing orbital energy when its orbit within the group is in
phase with the orbit of the group within the main halo. In turn, this
enhances the survival probability of low mass systems by placing them
on orbits where they spend extended periods in the periphery of the
main halo, outside the region where tidal fields may effectively strip
and disrupt them.

The combination of these two effects (energy gain and enhanced
survival likelihood) leads to a strong mass dependence on the orbital
properties of associated subhalos at $z=0$. This is illustrated in
Fig.~\ref{fig:RaRtVmax}, where we show the ratio of apocenter
(estimated at $z=0$) to turnaround distance as a function of subhalo
peak circular velocity. This figure shows clearly that the most
massive subhalos are found today in orbits with apocentric distances
much smaller than their turnaround: halos with $V_{\rm max}^{\rm
acc}\sim 0.4 \, V_{200}$ (which corresponds to roughly $M_{\rm
sub}^{\rm acc} \sim 0.1\, M_{200}$) have median apocenters of order
half their turnaround distance. On the other hand, the median
apocenter of associated subhalos with $V_{\rm max}^{\rm acc}\simless
0.1 V_{200}$ is of the order of the turnaround radius.

Note as well that the $V_{\rm max}$ dependence is quite pronounced at
the high-mass end but rather weak for low-mass subhalos. This
presumably reflects the fact that, once a subhalo is small enough, it
behaves more or less like a test particle in the potential of the main
system.

Finally, note that the mass dependence is less pronounced when the
{\it present-day} subhalo $V_{\rm max}$ is used. This is because tidal
stripping has a more pronounced effect on systems that orbit nearer
the center of the main halo. The more massive the subhalo at accretion
the closer to the center it is drawn and the more substantial its mass
loss, weakening the mass-dependent bias illustrated in
Fig.~\ref{fig:RaRtVmax}. We shall see below that the mass dependence
becomes even weaker when expressed in terms of the {\it present-day}
subhalo mass.

\begin{figure*} [ht]
\begin{center}
\includegraphics[width=\linewidth,clip]{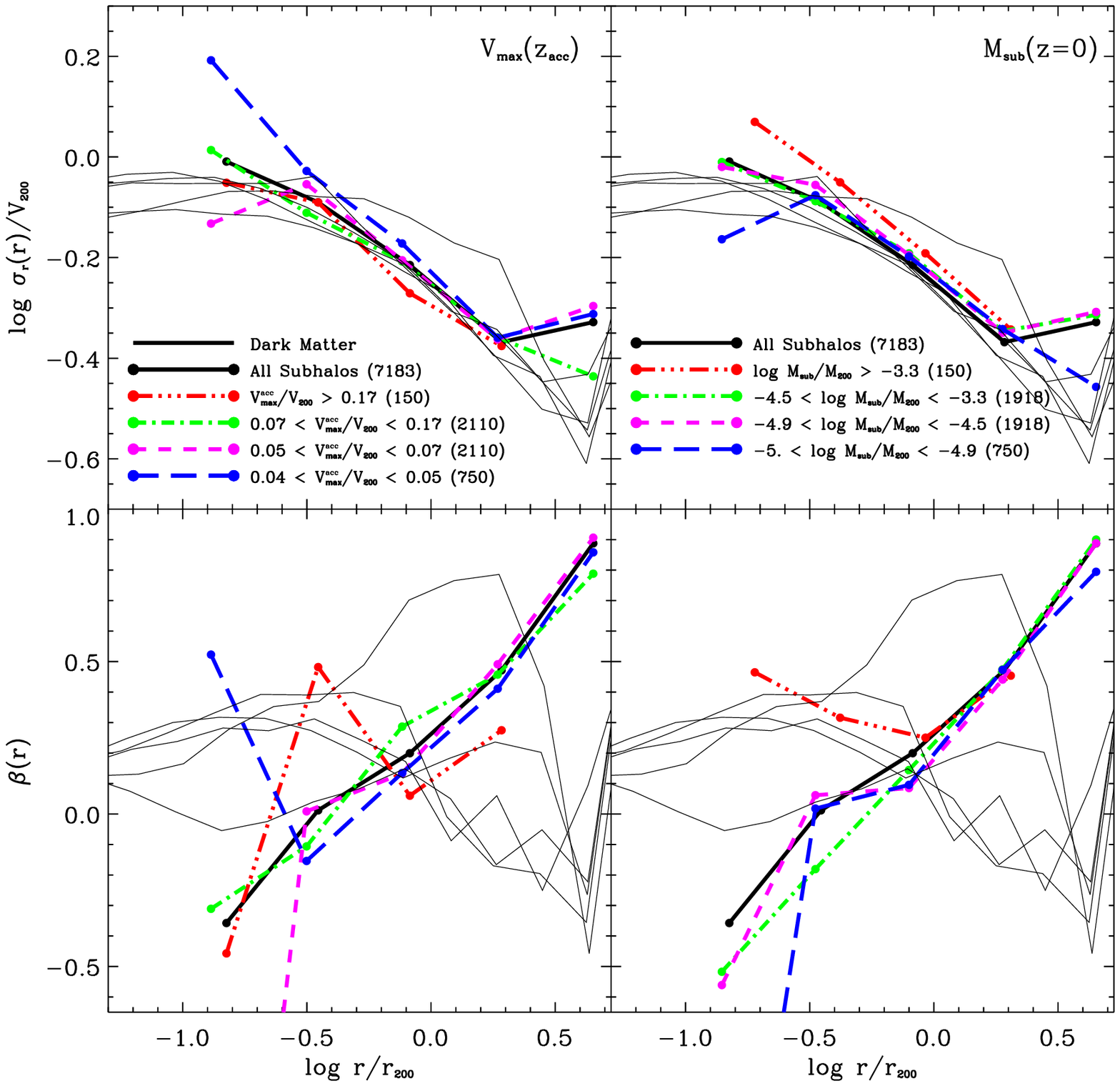}
\caption[] 
{\footnotesize{{Radial velocity dispersion and anisotropy profiles for
dark matter (thin black lines) and associated subhalos (thick colored
lines). Symbols are described in the legend and are the same as in
Fig.~\ref{fig:NrhoFit}. Note that the mass-dependent bias shown in
Fig.~\ref{fig:NrhoFit} is also reflected in the subhalo kinematics:
low mass subhalos tend to have higher velocity dispersions than their
high-mass counterparts. This bias is clearer when measuring subhalo
mass by the peak circular velocity at accretion time, $V_{\rm
max}^{\rm acc}$, rather than by the self-bound mass at $z=0$, $M_{\rm
sub}$. Note as well that subhalos tend to be on orbits less radially
biased than the dark matter, especially near the center. This is
presumably because subhalos on tangentially-biased orbits avoid the
innermost regions of the main halo, thus enhancing their survival
probability.}}}
\label{fig:VdispProf}
\end{center}
\end{figure*}

\subsection{Subhalo spatial distribution}
\label{ssec:subhrdist}

The number density profile of all associated subhalos is shown by the solid (black) curve in Fig.~\ref{fig:NrhoFit}. The profile may be approximated rather accurately by the same empirical formula introduced by Navarro et al (2004) to describe the mass profile of CDM halos. This profile is characterized by a power-law dependence on radius of the logarithmic slope of the density, $d\log\rho/d\log r \propto r^{\alpha}$, which implies a density profile of the form,
\begin{equation}
\ln (n(r)/n_{-2}) = -(2/\alpha) [(r/r_{-2})^\alpha -1].
\label{eq:rhoalpha}
\end{equation}
This density law was first introduced by Einasto (1965), who used it
to describe the distribution of old stars within the Milky Way. For
convenience, we will refer to it as the Einasto profile. The scaling
parameters $n_{-2}$ and $r_{-2}$ may also be expressed in terms of
the {\it central} value of the density, $n_0=n(r=0)=\exp (2/\alpha)\,
n_{-2}$, and of the radius containing {\it half} of the associated
subhalos, $r_{\rm h}$.

We list in Table~\ref{tab:nrho} the parameters obtained by fitting eq.~\ref{eq:rhoalpha} to the subhalo number density profiles. (Note that the units used for $n_0$ are arbitrary, but they are consistent, in a relative sense, for the various subhalo populations.) As discussed by Navarro et al (2004), Merritt et al (2005, 2006), and more recently by Gao et al (2007), $\Lambda$CDM halo density profiles are well described by $\alpha_{\rm DM}$ in the range $\sim 0.15-0.3$. This is in sharp contrast with the much larger values obtained for the subhalo number density profile ($\alpha_{\rm sub}\sim 1.0$; i.e., the 3D radial distribution of subhalos is approximately ``exponential''), and quantifies the well-established spatial bias between the subhalo population and the dark matter mass profile of the main halo. The larger values of $\alpha$ characterizing the subhalo spatial distribution imply a large nearly constant density ``core'' in their profile, in contrast with the ``cuspy'' density profile of the dark halo, shown as a solid line (without symbols) in Fig.~\ref{fig:NrhoFit}.

The left panel in Fig.~\ref{fig:NrhoFit} shows that the subhalo spatial distribution depends sensitively on subhalo mass, as measured by the peak circular velocity at accretion, $V_{\rm max}^{\rm acc}$ (see also, Nagai \& Kravtsov (2005), Faltenbacher \& Diemand (2006), Kuhlen et al (2007)). The various colored profiles in this panel correspond to splitting the sample of subhalos in four groups, according to the value of $V_{\rm max}^{\rm acc}$ (normalized to $V_{200}$, the virial velocity of the main halo at $z=0$). The concentration increases systematically with $V_{\rm max}^{\rm acc}$; for example, half of the $\sim 150$ (surviving) subhalos with $V_{\rm max}^{\rm acc} > 0.17\, V_{200}$ are contained within $\sim 0.7 \, r_{200}$ at $z=0$. The corresponding radius for subhalos with $0.04<V_{\rm max}^{\rm acc}/V_{200} < 0.05$ is $\sim 1.1 \, r_{200}$ (see details in Table~\ref{tab:nrho}).

Interestingly, the mass dependence of the subhalo number density profile essentially disappears when the {\it present-day} subhalo mass is used to split the sample. This is illustrated in the right-hand panels of Fig.~\ref{fig:NrhoFit}, which shows that the shape of the density profile of subhalos differing by up to two decades in mass is basically the {\it same}. This is in agreement with the earlier results of Gao et al (2004) and Diemand et al (2004), but indicates that the apparent mass-independence of the subhalo spatial distribution is {\it not} the result of efficient mixing within the main halo, but rather a somewhat fortuitous result of the cancellation of the prevailing trend by dynamical friction and tidal stripping.

It is conceivable that numerical artifact may also help to erase the dependence of $n_{\rm sub}(r)$ on present-day subhalo mass. Indeed, {\tt SUBFIND} (like every subhalo finder) will tend to assign masses to subhalos which depend slightly, but systematically, on their location within the main halo. The mass of subhalos near the center is more likely to be underestimated, and some subhalos may, indeed, even be missed altogether if close enough to the central cusp. Splitting the sample by $V_{\rm max}^{\rm acc}$ minimizes such effects and allows for the subhalo mass bias to be properly established.

\subsection{Velocity  anisotropies}
\label{ssec:subhvelan}

The mass dependence of the subhalo spatial distribution discussed in
the previous subsection is significant, but not very large, and thus
is less clearly apparent in their kinematics, as shown in
Fig.~\ref{fig:VdispProf}. The top panels of this figure show the
radial velocity dispersion profile, $\sigma_r=\langle v_r^2
\rangle^{1/2}$, computed in spherical shells for the same subsamples
discussed in Fig.~\ref{fig:NrhoFit}.  The bottom panels show the
anisotropy profile, defined as $\beta\equiv
1-(\sigma_{\theta}^2+\sigma_{\phi}^2)/2\sigma_r^2$. The mean values of
the velocity dispersion for each component are listed in
Table~\ref{tab:nrho}.

The solid lines without symbols in Fig.~\ref{fig:VdispProf} correspond to dark matter particles of the main halo, randomly sampled in order to match the total number of subhalos. As expected, the dark matter velocity distribution is mildly anisotropic, with a radial bias that increases outward and reaches a maximum near the virial radius.

The radial velocity dispersion profile of the subhalo population
follows closely that of the dark matter, although as a whole, the
subhalo population is kinematically biased relative to the dark
halo. The effect, however, is barely detectable; we find
$\sigma_r^{\rm sub}/\sigma_r^{\rm DM} \sim 0.98$.
Our results thus confirm the earlier conclusions of Ghigna et al
(1998), Gao et al (2004), Diemand et al (2004) about the presence of a
slight kinematic bias between subhalos and dark matter. 

Unlike the conclusions of Diemand et al, however, we find a
significant discrepancy between the anisotropy profiles of the subhalo
population and of the dark halo. As shown in the lower panels of
Fig.~\ref{fig:VdispProf}, subhalos are on orbits less dominated by
radial motions than the dark matter and, indeed, have a pronounced
{\it tangential} bias near the center (i.e., for $r\simless 0.3\,
r_{200}$).  With hindsight, this is not entirely unexpected, since
subhalos nearer the center are more likely to survive if they are on
tangentially-biased orbits that keep them away from the innermost
regions of the halo, where tidal effects are strongest.

\begin{figure*} [ht]
\begin{center}
\includegraphics[width=\linewidth,clip]{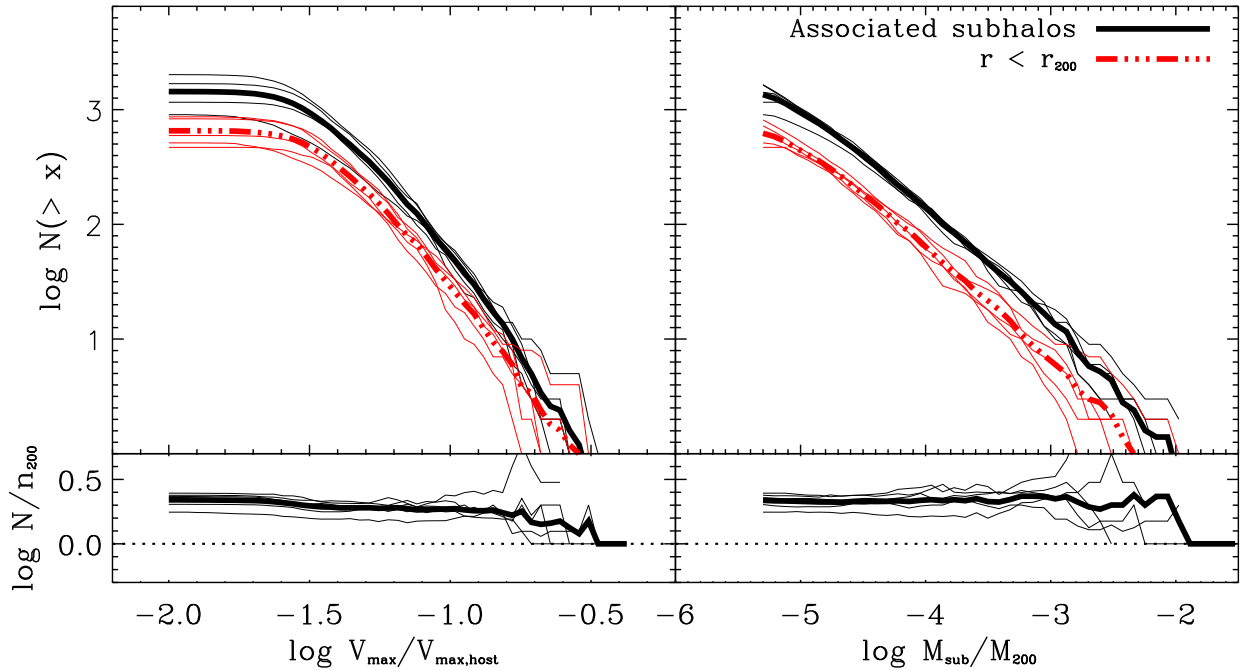}
\caption[] 
{\footnotesize{{Associated subhalo mass ($M_{\rm sub}$) and peak circular velocity ($V_{\rm max}$) cumulative distributions (both quantities measured at $z=0$). Black lines correspond to all associated subhalos; red lines to subhalos identified within $r_{200}$. Thick lines in each panel denote the average of our $5$ simulations. Note that the number of associated subhalos exceeds by about a factor of $\sim 2$ the number of subhalos found within $r_{200}$. The residuals are computed relative to the subhalo  population within the virial radius. }}}
\label{fig:MsVmf}
\end{center}
\end{figure*}

\subsection{Subhalo mass function}
\label{ssec:subhmf}

The large number of associated subhalos on high-energy orbits
discussed above imply that subhalos within the virial radius are just
a fraction of all subhalos physically influenced by the main
halo. This is illustrated quantitatively in Fig.~\ref{fig:MsVmf},
where we show the cumulative peak circular velocity and mass functions
of subhalos identified within $r_{200}$. The thin red lines in this
figure correspond to subhalos identified within $r_{200}$; black to
the full sample of associated subhalos.  Thick lines show the average
results for the $5$ simulated halos considered here. The residuals
shown in the small panels are computed relative to the average for the
associated subhalo population, and show that, on average, the total
number of associated subhalos exceed those within $r_{200}$ by a
factor of $\sim 2$. 

Fig.~\ref{fig:MsVmf} illustrates a number of interesting results. One
is that, at the low mass end, the shape of the subhalo mass and
velocity function is insensitive to the radius adopted for
selection. Indeed, there is no obvious systematic trend with $V_{\rm
max}$ or $M_{\rm sub}$ for $V_{\rm max}/V_{\rm max}^{\rm host}
\simless 0.2$.  Below certain threshold, low mass subhalos behave as
``test particles'' in the potential of the main halo and their radial
distribution becomes independent of mass. This implies that attempts
to determine the asymptotic slope of the subhalo mass function are
unlikely to be compromised by selecting for analysis only halos within
the virial radius, as is traditionally done.

On the other hand, the subhalo mass function {\it shape} is
substantially affected at the opposite end; although about half of all
associated subhalos with $V_{\rm max} \simless 0.15 \, V_{\rm
max}^{\rm host}$ are missing from within $r_{200}$, this fraction
declines to one quarter for $V_{\rm max} \sim 0.28 \, V_{\rm max}^{\rm
host}$, and to zero for $V_{\rm max}> 0.31 \, V_{\rm max}^{\rm
host}$. As a result, in that mass range, the mass function of subhalos
identified within $r_{200}$ is shallower than that of associated
systems. This should have interesting consequences for semianalytic
modeling of the luminosity function in galaxy groups and clusters,
which traditionally assume that all accreted subhalos remain within
the virial radius of the main system.

\section{Summary and Discussion}
\label{sec:conc}

We have used a suite of cosmological N-body simulations to study the
orbital properties of substructure halos (subhalos) in galaxy-sized
cold dark matter halos. We extend prior work on the subject by
considering the whole population of {\it associated} subhalos, defined
as those that (i) survive as self-bound entities to $z=0$, and (ii)
have at some time in the past been within the virial radius of the
halo's main progenitor. Our main findings may be summarized as
follows.
\begin{itemize}

\item 
The population of {\it associated} subhalos extends well beyond three
times the virial radius, $r_{200}$, and contains a number of objects
on extreme orbits, including a few with velocities approaching the
nominal escape speed from the system. These are typically the low-mass
members of accreted groups which are propelled onto high energy orbits
during the tidal dissociation of the group in the potential of the
main halo.

\item 
The net result of this effect is to push low-mass subhalos to the periphery of the system, creating a well-defined mass-dependent bias in the spatial distribution of associated subhalos. For example, only about $\sim 29\%$ of subhalos which, at accretion time, had peak circular velocities of order $3\%$ of the present-day virial velocity ($V_{\rm max}^{\rm acc} \sim 0.03 \, V_{200}$), are found today within $r_{200}$.  This fraction climbs to $\sim 61\%$ and to $\sim 78\%$ for subhalos with $V_{\rm max}^{\rm acc}\sim 0.1\, V_{200}$ and $\sim 0.3 \, V_{200}$, respectively.

\item
The strength of the bias is much weaker when expressed in terms of the
subhalo {\it present-day} mass, due to the increased effect of
dynamical friction and tidal stripping on the most massive subsystems.

\item
The spatial distribution, kinematics, and velocity anisotropy of the
subhalo population are distinct from the properties of the dark
matter. Subhalos are less centrally concentrated, have a mild velocity
bias, and are, near the center, on more tangential orbits than the
dark matter.

\end{itemize}

The unorthodox orbits of substructure halos that result from the complex history of accretion in hierarchical formation scenarios have a number of interesting implications for theoretical and observational studies of substructure and of the general halo population.

One implication is that subhalos identified within the virial radius represent a rather incomplete census of the substructures physically related to (and affected by) a massive halo. This affects, for example, the interpretation of galaxy properties in the periphery of galaxy clusters, and confirms earlier suggestions that evolutionary effects normally associated with passage through the innermost regions of a massive halo, such as tidal truncation or ram-pressure stripping, should be detectable well outside the traditional virial boundaries of a group or cluster (Balogh, Navarro \& Norris 2000).

Furthermore, associated subhalos pushed well outside the virial radius of their main halo might be erroneously identified as separate, isolated structures in studies that do not follow in detail the orbital trajectories of each system. This effect would be most prevalent at low masses, and it is likely to have a significant effect on the internal properties of halos in the vicinity of massive systems. We expect, for example, halos in the periphery of groups/clusters to show evidence of truncation and stripping, such as higher concentrations and/or sharp cutoffs in their outer mass profiles.

The same effect may also introduce a substantial environmental
dependence in the formation-time dependence of halo clustering
reported in recent studies (Gao et al 2005; Zhu et al 2006; Jing et al
2007; see also Diemand et al 2007b). In particular, at fixed
mass, early collapsing halos might be more clustered because they are
physically associated with a more massive system from which they were
expelled.

A proposal along these lines has recently been advanced by Wang, Mo \&
Jing (2007) (see also Hahn et al 2008), who argue that such
environmental effects might be fully responsible for the
age-dependence of halo clustering. Our physical interpretation,
however, differs in detail from theirs. Whereas Wang et al argue for
the suppression of mass accretion onto ``old'' halos by ``heating by
large-scale tidal fields'' as responsible for their enhanced
clustering, our results suggest that the real culprit is the orbital
energy gain associated with the tidal dissociation of bound groups of
subhalos, which allows ``old'' low-mass halos to evade merging and to
survive in the vicinity of massive systems until the present.

A further implication of our results concern the spatial bias of the
most massive substructures discussed in S.~\ref{ssec:subhrdist}. If,
for example, luminous substructures in the Local Group trace the most
massive associated subhalos at the time of accretion, they may
actually be significantly more concentrated and kinematically biased
relative to the dark matter, a result that ought to be taken into
account when using satellite dynamics to place constraints on the mass
of the halos of the Milky Way and M31.

Finally, as already pointed out by Sales et al (2007a,b),
gravitational interactions during accretion may also be responsible
for the presence of dynamical outliers in the Local Group, such as Leo
I and And XII. Further work is needed to assess whether the
exceptional orbits of such systems could indeed have originated in the
tidal dissociation of groups recently accreted into the Local
Group. Since the latest proper motion studies of the Magellanic Clouds
seem to suggest that the Clouds are on their first pericentric passage
(Kallivayalil et al 2006; Piatek et al 2007), this is a possibility to
consider seriously when trying to puzzle out the significance of the
motion of the satellites of the Local Group.

\vskip 2.5cm 
We thank Simon White and Vincent Eke for useful discussions, and
an anonymous referee for a constructive report. ADL would like to
thank Jorge Pe\~{n}arrubia and Scott Chapman for many useful
discussion which have improved this work. The simulations reported
here were run on the Llaima Cluster at the University of Victoria, and
on the McKenzie Cluster at the Canadian Institute for Theoretical
Astrophysics. This work has been supported by various grants to JFN
and a post-graduate scholarship to ADL from Canada's NSERC. AH
gratefully acknowledges financial support from NOVA and NWO.


\begin{table*}
\center
\caption{Properties of simulated halos used in this study.}
\begin{tabular}{l c c c c c  c c c c c c c c } \hline \hline
Halo & $\epsilon_{\rm G}$& $M_{200}$ & $M_{\rm DM}^{\rm assoc}$ &$r_{200}$ & $V_{\rm max}$ & $r_{\rm max}$ & $N_{200}$ &$M_{\rm sub}^{\rm assoc}$ & $N_{\rm sub}$ & $N_{\rm sub}$ & $N_{\rm sub}$ &
$N_{\rm sub}$ & $N_{\rm sub}$ \\
     &  [kpc/h]    & [$M_{\odot}$/h]&[$M_{\odot}$/h] & [kpc/h] & [km/s] & [kpc/h] &          & [$M_{\odot}$/h] &     [assoc]    &($r<r_{200}$)  & ($r<r_{100}$) & ($r<r_{50}$)   & $r_{\rm apo}>2.5r_{\rm ta}$\\
\hline
9-1-53     &   0.250     &   9.17$\times10^{11}$& 12.0$\times10^{11}$&  158.0 &  184.8 & 36.0 &   3.24e6       & 0.89$\times10^{11}$& 904       &    513   &  742 & 1017 &    3     \\
9-12-46    &   0.181     &   6.44$\times10^{11}$& 9.79$\times10^{11}$& 140.4  &  159.9 & 34.9 &   4.82e6       & 0.47$\times10^{11}$ & 2020      &    865   &  1314& 1828 &     14    \\
9-13-74    &   0.220     &   8.76$\times10^{11}$& 12.4$\times10^{11}$& 155.6   &  175.9 & 32.0 &   4.18e6       & 1.04$\times10^{11}$&  1683      &    831   &  1232& 1645 &   15    \\
9-14-39    &   0.186     &   12.6$\times10^{11}$& 17.9$\times10^{11}$& 175.7  &  203.5 & 37.8 &   3.30e6       & 1.27$\times10^{11}$ & 1416      &    594   &  1050& 1581 &   4     \\
9-14-56    &   0.275     &   8.57$\times10^{11}$& 12.2$\times10^{11}$& 154.4   &  178.6 & 33.7 &   2.61e6       &1.00$\times10^{11}$ &  1160      &    469   &  646 & 930  &   12    \\
\hline
\end{tabular}\label{tab:numexp}
\end{table*}
\begin{table*}
\center
\caption{Dynamical properties of subhalo population}
\begin{tabular}{c| c c c c c c c c}\hline 
Sample                                   &$N_{\rm sub}$      &  $n_{0}$    & $r_{-2}$ &$r_{\rm h}$&$\alpha$&$\langle \beta \rangle$&$\langle \sigma_r \rangle$ \\ 
                                         &[5 sims] & [arb.units] &$[r_{200}]$&$[r_{200}]$&        &                       & $      [V_{200}]$          \\\hline\hline
Assoc. dark matter                       &         &   5422      &  0.l12   &  0.558    & 0.159  &  0.270                &        0.734              \\
                                         &         &             &          &           &        &                                                  \\
Assoc. subhalos                          & 7183    & 4691.0      &   1.08   &  1.09    & 0.842  &         0.053         &  0.712                    \\
                                         &         &             &          &           &        &                                                  \\
($X=V_{\rm max}^{\rm acc}/V_{200}$)  &             &        &          &               &           &            &                                    \\
${\rm X}> 0.170$                         &  150    & 42848.5      & 0.70     & 0.70      &  0.352 &         0.059          &  0.699                \\
$0.063 < X < 0.17$                       & 2110    & 3709.2       & 0.80     & 0.87     &   0.754 &         0.091         &  0.712                \\
$0.040 < X < 0.063 $                     & 2110    & 875.9        & 1.09     & 1.10     &  0.991 &         0.018         &  0.699                 \\
$0.025 < X < 0.040 $                     &  750    & 222.0        & 1.26     & 1.39     &   1.001 &         0.034         &  0.762                \\
                                         &         &              &          &           &         &                                             \\
($Y=\log_{10}M_{\rm sub}/M_{200}$)  &         &        &          &          &           &                                                        \\
$Y> -3.3  $                              &  150    & 28.3         & 1.12     &  1.14     &  1.536 &         0.207         &  0.769                 \\
$-4.6  < Y < -3.3$                       & 1918    & 1181.3       & 0.99     &  1.06     &  0.886 &         -0.001        &  0.723               \\
$-5.1  < Y < -4.6$                       & 1918    & 1275.5       & 1.11     &  1.08     &  0.897 &         0.021         &  0.735               \\
$-5.42 < Y < -5.1$                       &  750    & 341.6        & 1.09     &  1.13     &  0.939 &         -0.028        &  0.708             \\
\end{tabular}\label{tab:nrho}
\end{table*}

{}

\end{document}